\documentclass[12pt,preprint]{aastex}

\shorttitle{The outskirts of the SMC}
\shortauthors{No\"el et al.}

\begin{document}


\title{Stellar Populations in the Outskirts of 
the Small Magellanic Cloud: No Outer Edge Yet}

\author{Noelia E. D. No\"el\altaffilmark{1} and Carme Gallart\altaffilmark{1}}


\altaffiltext{1}{Instituto de Astrof\'\i sica de 
Canarias, 38200, La Laguna, Tenerife, Spain; noelia@iac.es}

\begin{abstract}

We report the detection of intermediate-age and old stars belonging to the SMC at 6.5 kpc from the SMC center in the southern direction. 
We show, from the analysis of three high quality 34\arcmin $\times$ 33\arcmin CMDs,
 that the age composition of the stellar population is similar at  galactocentric distances of $\thicksim$4.7 kpc, $\thicksim$5.6 kpc, and $\thicksim$6.5 kpc. 
 The surface brightness profile of the SMC follows an exponential law, with no evidence of truncation, all the way out to 6.5 kpc. 
 These results, taken together, suggest that the SMC `disk' population is dominating over a possible old Milky Way-like stellar halo,
  and that
 the SMC may be significantly larger than previously thought.  
 
\end{abstract}




\section{INTRODUCTION} \label{intro}

Many important clues about the galaxy assembly process lie in the faint outskirts of galaxies. 
For example, recent galaxy formation simulations suggest that almost all galaxies, even the smallest, should contain an
extended old, metal poor, stellar halo (see e.g. Bekki \& Chiba 2005; Read et al. 2006a). Furthermore, evidence of recent merging and/or interaction should remain in the form of extended tidal debris 
(see e.g. Mu\~noz et al. 2006). The distribution and extent of such debris, or a characteristic
 underlying old stellar halo are both sensitive to the total mass and extent of a galaxy. Thus, probing the faint edges of galaxies can also give us constraints on their dark matter content. 

The extended extreme edges of galaxies are observed to be so faint ($\mu$$\geq$28 mag/arcsec$^{2}$; see e.g. Gallart et al. 2004) 
that their study has been limited so far to just the Local Group dwarf galaxies, in which we can study the flux from the individual stars.
 Such studies began with the discussion by Sandage (1962) on the outer parts of IC 1613, which showed
 faint red stars significantly extended beyond the galaxy's central irregular body. More recently, evidence for outer, faint stellar envelopes
  has been found, for example, in WLM 
 (Minniti \& Zijlstra 1996) and Leo A (Vansevi$\check{c}$ius et al. 2004). However, it may be premature
  to label these extended stellar distributions as an `old stellar halo' like that observed in the Milky Way. There is evidence to suggest 
  that the faint outer stellar populations of nearby dwarf irregulars  are composed of a range of ages 
  (Old main sequence [MS] turnoff photometry: Gallart et al. 2004, Hidalgo et al. 2003; C stars: Letarte et al. 2002; 
  CMD modelling: Aparicio \& Tikhonov 2000). In some cases, the
   extended stellar light is almost certainly tidal debris (e.g. the case of Carina, Mu\~noz et al. 2006). 

In this letter, we present the first evidence for an extended distribution of stars in the southern direction of the 
Small Magellanic Cloud (SMC), 
up to $\thicksim$6.5 kpc (5.8\degr) from the optical centre of the galaxy. We 
provide unambiguous information about the age composition of this population from a CMD reaching the oldest MS turnoffs. 
 Photometric studies of the outer SMC began with Gardiner \&
Hatzidimitriou (1992) who studied the age composition with CMDs reaching a magnitude limit at R=20, and hence, with 
 MS stars for which it was possible to determine precise ages up to $\thicksim$2 Gyr ago.  
Their contour plots of the surface distribution of horizontal branch (HB)/clump stars reached a galactocentric radius of $\thicksim$6 kpc 
 in the semi-major axis direction (toward the LMC) but the star counts in these outer parts
are within the noise level.
  Carbon stars were also detected in the outer regions of the SMC (Hatzidimitriou et al. 1997; Kunkel et al.
 2000). Few of these intermediate-age tracers were found in the southern direction up to $\thicksim$7 kpc. 
 Demers \& Battinelli (1998) observed 5 fields in the outer SMC Wing and in the Magellanic Bridge, finding a burst of star formation
  occurred
between 10 and 25 Myr ago. Harris (2007) presented a detailed analysis of the star formation history of the young inter-Cloud population
along the ridgeline of the HI gas that forms the Magellanic Bridge and searched for the older population in this area. He found  
an intermediate-age and old population of at 4.4\degr and 4.9\degr from the SMC center in that direction but only young
population at 6.4\degr. 
  Recently, No\"el et al. (2007) presented the age distribution of 12 SMC fields,
    through CMDs which reach the oldest MS turnoff with an excellent photometric accuracy.
     From the CMD of their outermost field, they inferred that, at $\thicksim$4 kpc from the center, the intermediate-age population 
       is still substantial. 
      Here, we extend this analysis using images obtained with the WFI at the ESO 2.2m telescope in La Silla,
       Chile. We present the surface brightness together with the age composition of 3 fields in the southern direction of the SMC,
        at 4.2\degr ($\thicksim$4.7 kpc),  4.9\degr ($\thicksim$5.6 kpc), and 5.8\degr ($\thicksim$6.5 kpc)
	 from the SMC optical center, through the analysis of the CMDs of each of these fields.  

\section{OBSERVATIONS AND DATA REDUCTION} \label{obs}

In order to characterise the stellar content of the outer SMC and to search for the SMC outer edge, 
we obtained {\it B}- and {\it R}-band images of 3 SMC fields centered at 4.2\degr, 4.9\degr, and 5.8\degr from the SMC center, 
respectively. The observations were taken from June 2001 to June 2006, using the WFI attached to the 2.2m telescope 
at La Silla Observatory, Chile, with a 4$\times$2 mosaic of 2048$\times$4096 CCD's. Each chip covers a sky 
area of 8\arcmin.1 $\times$ 16\arcmin.2. This combination gives a field size of 34\arcmin $\times$ 33\arcmin 
(aproximately 0.64 kpc $\times$ 0.615 kpc) with a scale of 0.238\arcsec/pixel. 
Figure~\ref{SMC} shows the spatial distribution of our SMC fields. The large squares represent the WFI fields we present in this paper;
 they are far from the optical center of the SMC, and lie well-away from any observed HI gas 
(Stanimirovi$\check{c}$ et al. 1999). The small simbols denote the fields observed using the 100-inch telescope at LCO, presented in
No\"el et al. 2007. 

  The coordinates of the WFI fields and the data obtained for each of
 them are detailed in Table~\ref{fields}. The first column denotes the field name; the second and third columns the
 right ascension and the declination, respectively; the
  fourth column, the galactocentric distance
 (r); and the fifth and sixth columns the integration 
 times in {\it B} and {\it R}. Seeing was typically
between 0.$\arcsec$7 and 1.$\arcsec$2.

 For the basic reduction, bias exposures, sky flats, and dome flats were taken every night. 
  The reduction was performed using the MSCRED package within IRAF\footnote{IRAF is distributed by National Optical
   Astronomy Observatory, which is operated
     by the Association of Universities for Research in Astronomy, Inc., under cooperative agreement with National
      Science foundation.}.
   Profile-fitting photometry of the SMC fields was obtained using the DAOPHOT and ALLSTAR/ALLFRAME software packages (Stetson 1987, Stetson 1994). 
 We photometered every chip separately as in No\"el et al. (2007). 
 To compensate for the low density of stellar objects in the analyzed fields, we combined all chips to obtain a single deep CMD for each field.  
 The aperture corrections were obtained from synthetic aperture photometry by measuring several isolated bright stars through a 
series of increasing apertures and the
  construction of growth curves (Stetson 1990).
 Errors in the aperture corrections were calculated as the
  standard error of the mean of these differences, and were typically between $\pm$0.001 and $\pm$0.003. 
   During the photometric nights, Landolt (1992)                                                   
standard star fields SA92, SA95, SA104, SA110, SA113, RU149 were observed several
times for calibration purposes. 
  The total zero-point errors of the photometry, including the error in the extinction, in the aperture corrections, and the uncertainties in the calibrations, were $\sim$0.03 mag in both {\it B} and {\it R} bands. 

\section{The CMDs of the SMC outskirts} \label{cmd}

\begin{figure}
\includegraphics[width=5cm,height=5cm]{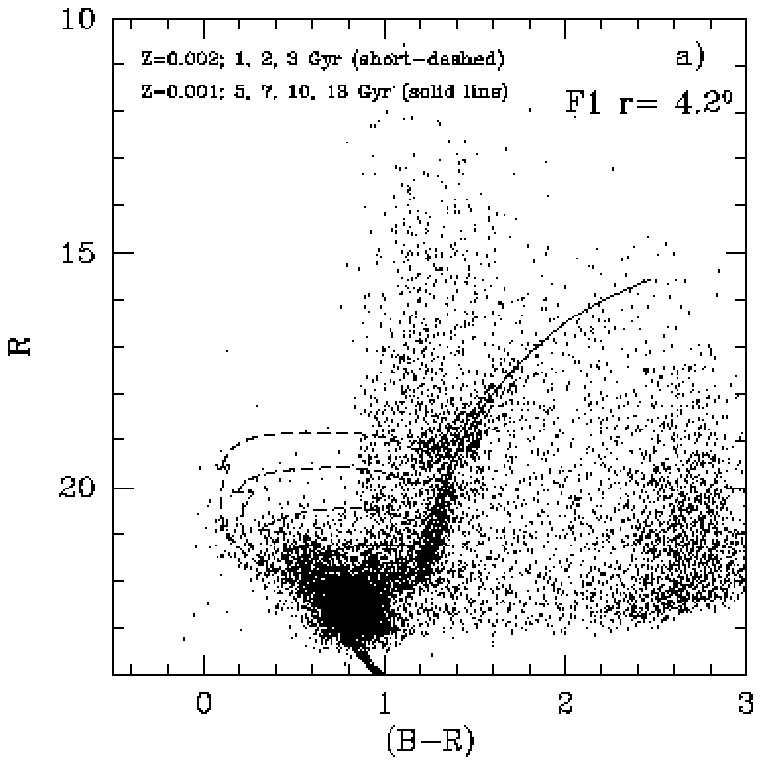}
\includegraphics[width=5cm,height=5cm]{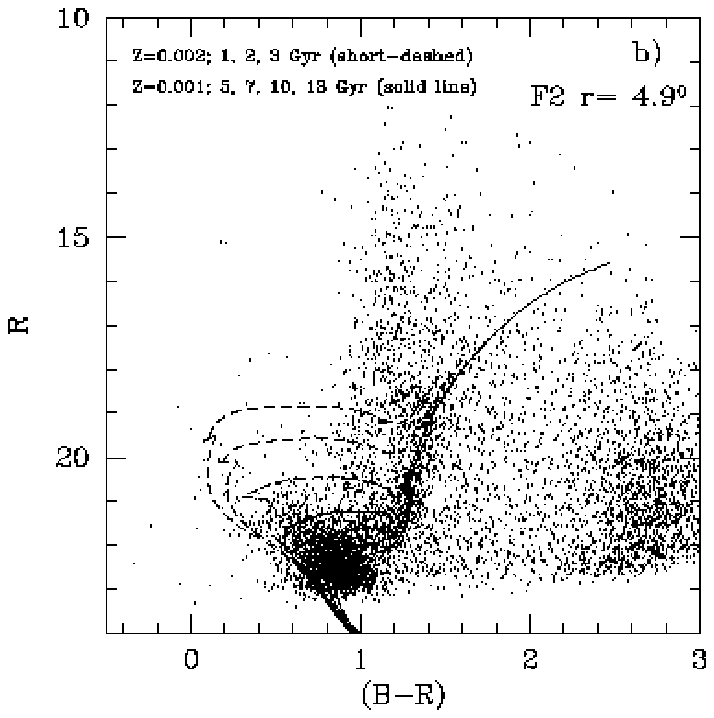}
\includegraphics[width=5cm,height=5cm]{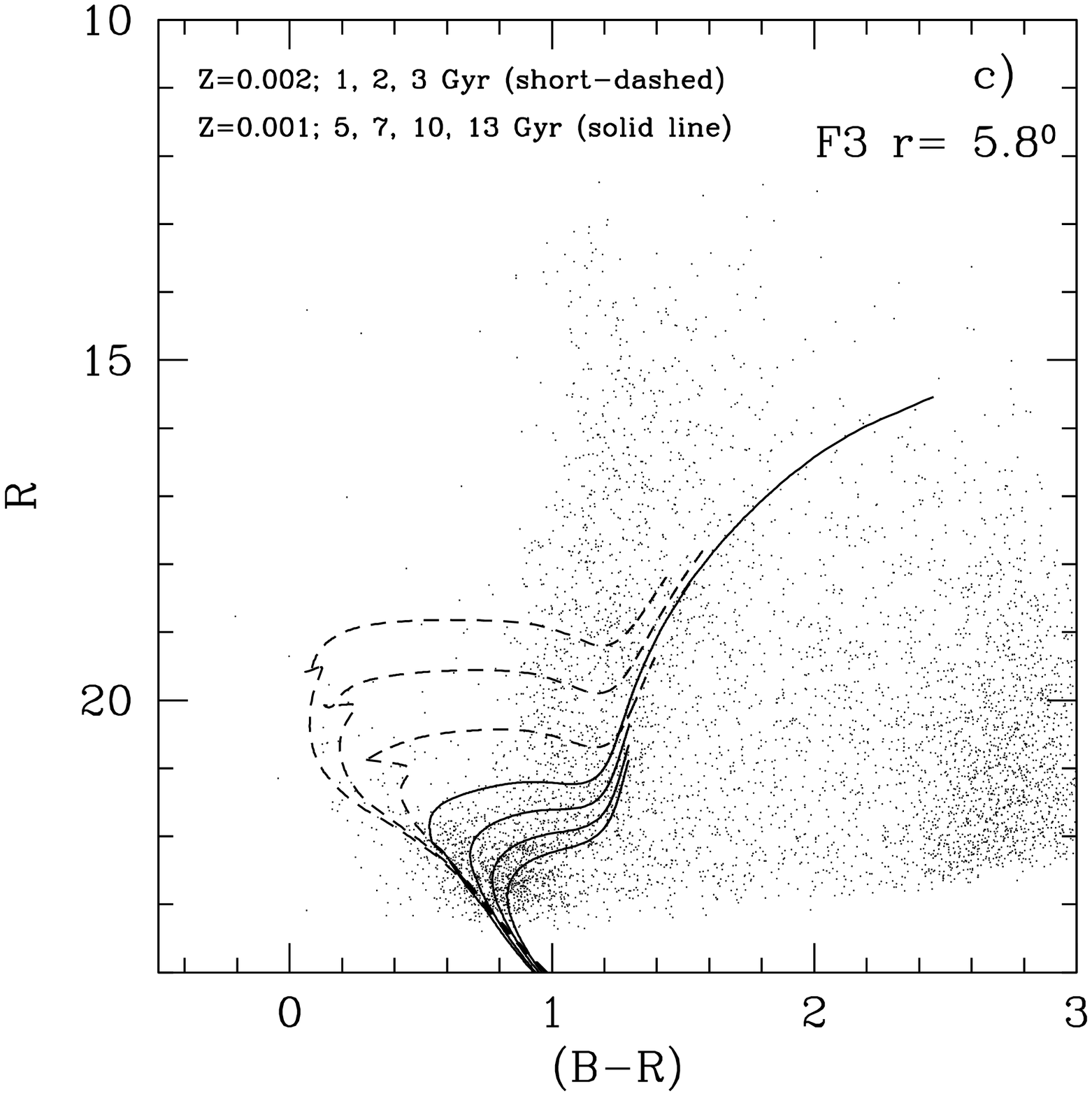}
\includegraphics[width=5cm,height=5cm]{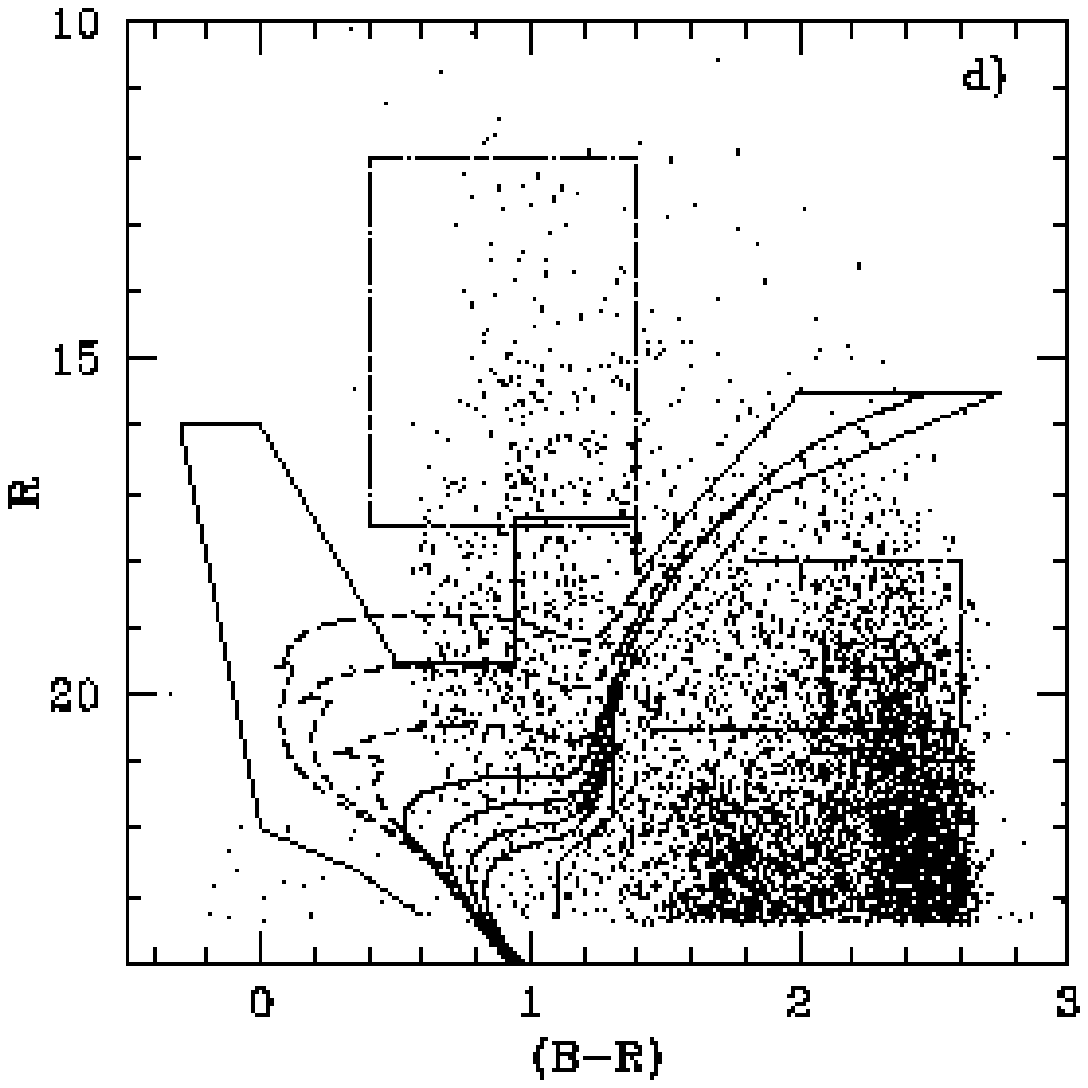}
\caption{ Panels a) to c) show the CMDs of our outer SMC fields in order of increasing distance from
the center of the galaxy. From left to right, the galactocentric distances are: 4.7 kpc, 5.6 kpc, and 6.5 kpc.
 BaSTI isochrones (Pietrinferni et al. 2004) were overlapped. Only the 5 Gyr isochrone is drawn up to 
 the upper red giant branch (RGB) in order to show the extension in color and magnitude 
of the RGB stars. The expected position of the RC is [{\it B,(B-R)}]=[18.9,1.3] and indeed an enhancement in the number of stars
is observed in fields F1 and F2 around this position.
Stars as young as 3 Gyr old are still present in fields F1 and F2, while there are a
non-negligible amount of stars around the 5 Gyr isochrone in our most peripheral field, F3. 
Panel d) shows the TRILEGAL simulation of the foreground stars 
(Girardi et al. 2005) in field F2. The dot-dashed  boxes indicates the portion of CMD devoided of SMC stars taken for the scaling 
(see text for details). BaSTI isochrones were overlapped in this simulation to show the
 loci of the SMC stars. The thick solid lines denote the areas of the CMD selected to obtain the SMC surface brightness.}
\label{three_outer}
\end{figure}

 Figures~\ref{three_outer}a, \ref{three_outer}b, and \ref{three_outer}c show  the $[{\it (B-R),R}]$ CMDs of the three SMC fields in order of increasing distance from the SMC center. BaSTI isochrones (Pietrinferni et al. 2004) are 
 overlapped (see caption for details). Stars with at least one valid measurement in each filter have been selected using the
following limits for the error and shape parameters given by ALLFRAME: $\sigma_{(B-R)} (= \sqrt{\sigma_B^2+\sigma_R^2}) \le 0.15$,
 $\mid sharp \mid \le 0.6$ and $\chi \le 5$.  A total of
 10230 
  stars down to $R\leq23$ were measured in field F1 
  (panel a in figure~\ref{three_outer}),  
  7685 
   in field F2 (panel b), and 
   5642 
   in field F3 (panel c).

The CMD of the innermost field, F1, shown in figure~\ref{three_outer}a, reaches 1.2 magnitudes below the old MS turnoff,
 while the CMDs of field F2 (figure~\ref{three_outer}b) and field F3 (figure~\ref{three_outer}c) reach 1 magnitude below this point. 
  Each of these CMDs shows, for the first time, the details of the age structure of the stellar population at these outer parts of the SMC. 
At 4.7 kpc and 5.6 kpc from the SMC center (figures~\ref{three_outer}a and ~\ref{three_outer}b, respectively),
 the areas around the 3 Gyr isochrone are quite well populated.
      It is noticeable that, at 6.5 kpc from the SMC center, there is still galaxy, with intermediate-age and old stars, 
 as seen in figure~\ref{three_outer}c. Most of the main features of a CMD are present in F3: 
 a quite-well populated intermediate-age
 and old MS and a well-defined subgiant branch. 
 The areas around the 5, 7, 10, and 13 Gyr isochrones are well populated in all of the CMDs and there is a lack of a blue extended HB in all of them. 
 
Panel  d) in figure~\ref{three_outer} shows 
the $[{\it (B-R),R}]$ CMD of the simulated 
 foreground Milky Way stars in the direction of our SMC field\footnote{We show only one CMD since
  the three fields have similar foreground stars distributions.} F2, as 
 obtained using the TRILEGAL code (Girardi et al. 2005). 
 In the simulation we used the (l,b) coordinates of each of our SMC fields, 
 an area in the sky of 34\arcmin $\times$ 33\arcmin, down to a magnitude of 23.4 in $R$, an IMF from 
 Kroupa (2001) corrected for binaries, dA$_{V}$/dr$_{\odot}$= 0.00015 mag/pc, thin and thick discs
 (squared hyperbolic secants) and a halo. 
 BaSTI isochrones were also overlapped in order to show the loci of the stars in the SMC fields.
 When the CMDs of our SMC fields (figures~\ref{three_outer}a to~\ref{three_outer}c) are compared with the foreground simulations,
  it is clear that the main features of the CMDs
mentioned above are not presented in the latter. There are only a few stars in the region occupied by 
the few Gyr old MS. There is a gradient in the population of predicted foreground stars in the range {\it (B-R)}$\leq$0.8, which increases  
while going further south. 

\section{The surface brightness profile: reaching the SMC outer edge?} \label {surfacebrightness}

We have derived the SMC surface brightness in the three observed fields. In order to minimize the error in the 
 substraction of the foreground stars' contamination,
 we integrated the flux only in the areas of the CMDs
 in which we expect SMC stars, as shown in figure~\ref{three_outer}d (thick solid lines). 
 We used the predictions of the TRILEGAL model described above  
 for foreground substraction purposes, 
but scaling the flux predicted for the Milky Way stars inside the fiducial SMC area shown in figure~\ref{three_outer}d 
 by the ratio of the observed and predicted flux in regions of the  CMDs 
 clearly devoided of SMC stars (dot-dashed boxes in figure~\ref{three_outer}d).

 In figure~\ref{surface_brightness}, we present the surface brightness (in mag/arcsec$^2$) of our fields
measured in ${\it B}$ (panel a)) and ${\it R}$ (panel b)) bands (filled circles), as a function of distance from the optical
center of the SMC.
 The surface brightness of the fields from No\"el et al. (2007) are also plotted. 
  The open triangles represent the eastern
 fiels, while the filled triangles are the western fields; the squares denote the southern fields. 
 
The total error bar is the difference between the values of the foreground flux in the fiducial SMC area 
as predicted by the TRILEGAL code and after scaling by the
ratio of observed and predicted fluxes.  The foreground flux predicted by TRILEGAL and actually observed in the boxes devoided 
of SMC stars in field F1 differed by about 50\% in both {\it B} and {\it R} bands in all fields.
 
 The measured surface brightness profile is well-fit by an exponential law, 
 with: $B(0)$=22.3$\pm$0.3 
 and $\alpha$$^{-1}$=47.62\arcmin$\pm$0.02, and  $R(0)$=21.5$\pm$0.2 and $\alpha$$^{-1}$=52.63\arcmin$\pm$0.01.
 Note the slightly shallower profile in {\it R}, reflecting the larger extent of the intermediate-age and old populations.

\section{Discussion}

\begin{figure}
\plotone{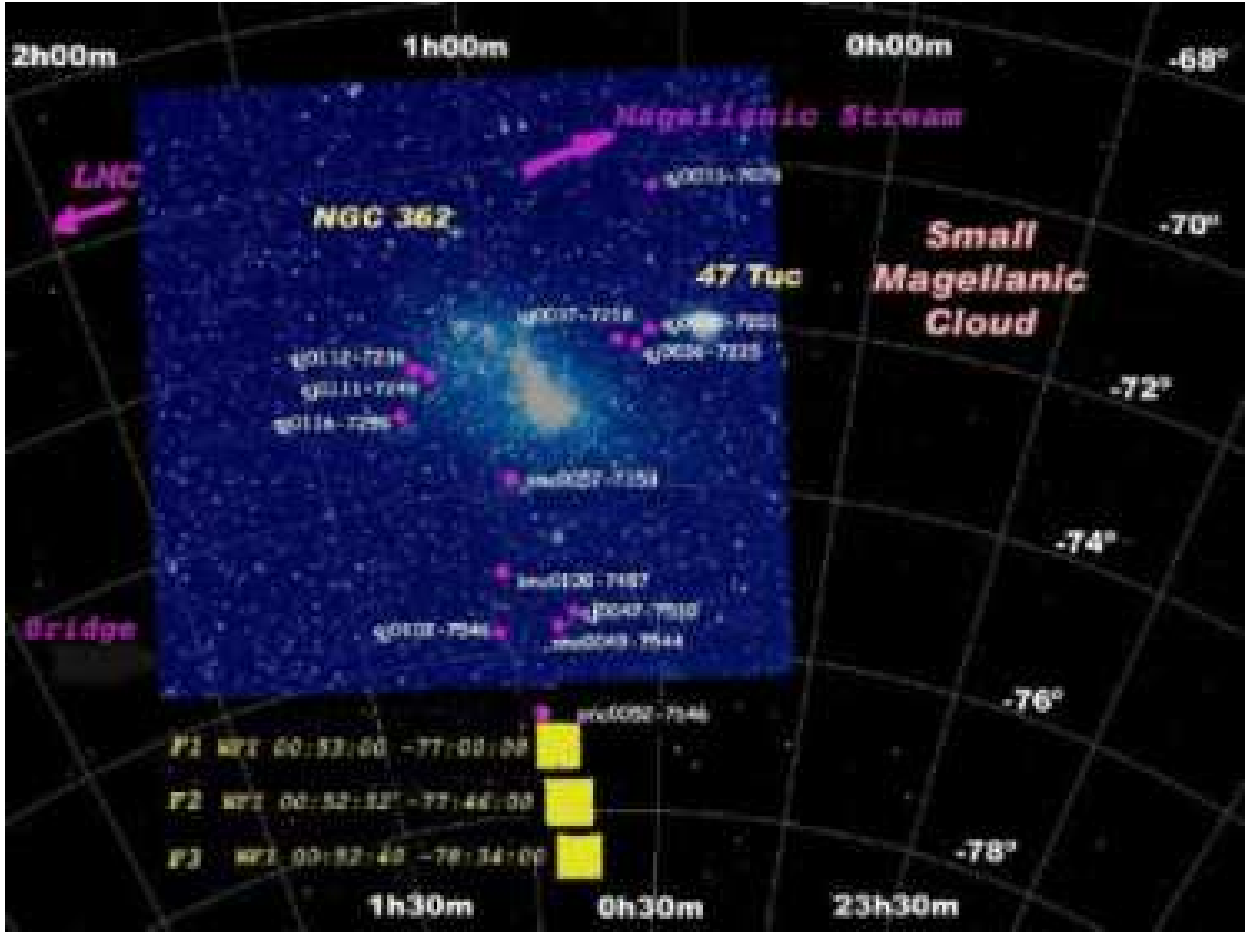}
\caption{Spatial distribution of our SMC fields. The large squares denote the 34$\arcmin$ $\times$ 33$\arcmin$ 
fields analyzed here. These WFI fields were chosen to be well away from any observed HI (Stanimirovi$\check{c}$ et al. 1999).
 The small simbols represent the fields analyzed in No\"el et al. 2007.\label{SMC}}
\end{figure}

\begin{figure}
\includegraphics[width=7cm,height=7cm]{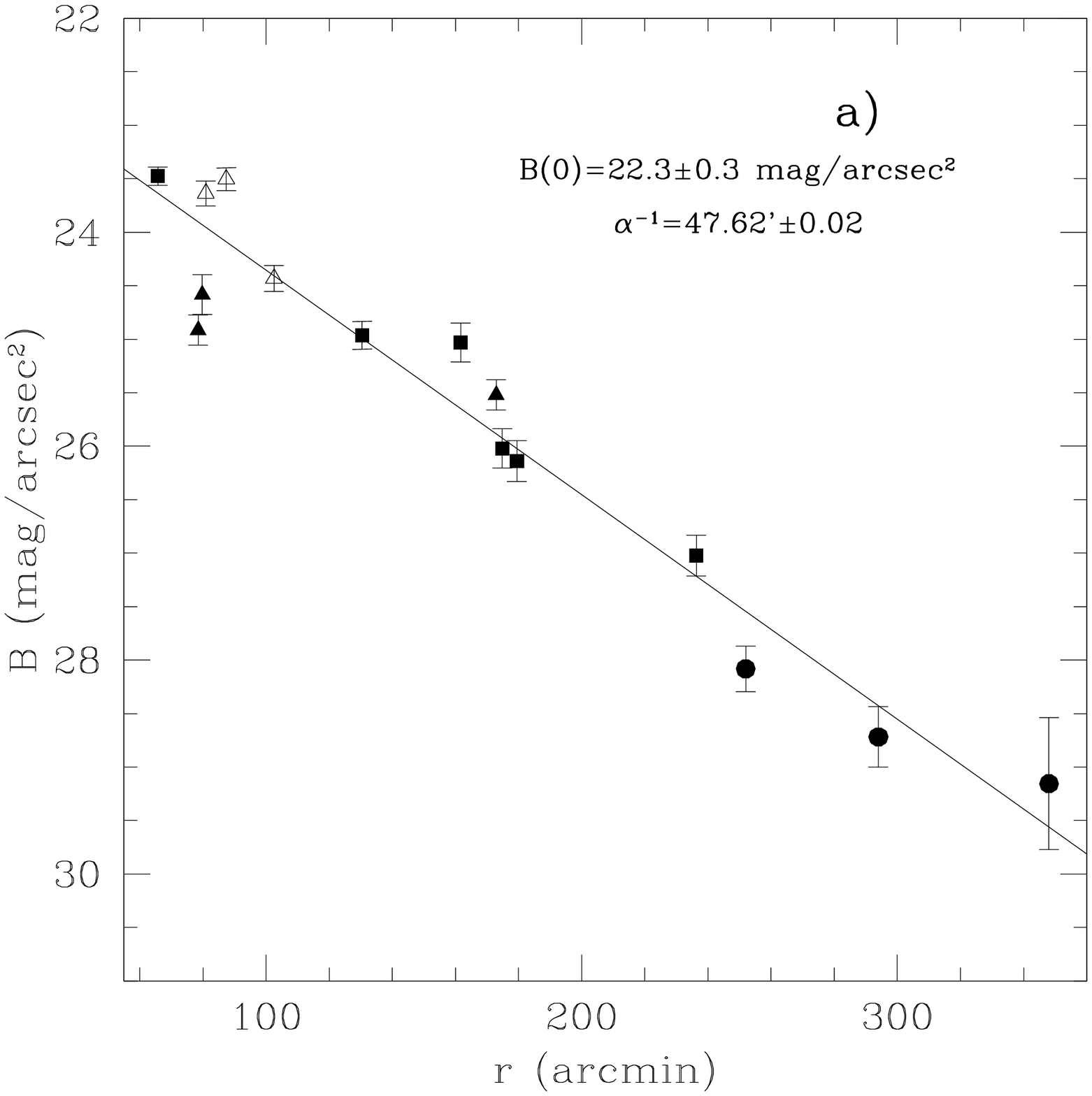}
\includegraphics[width=7cm,height=7cm]{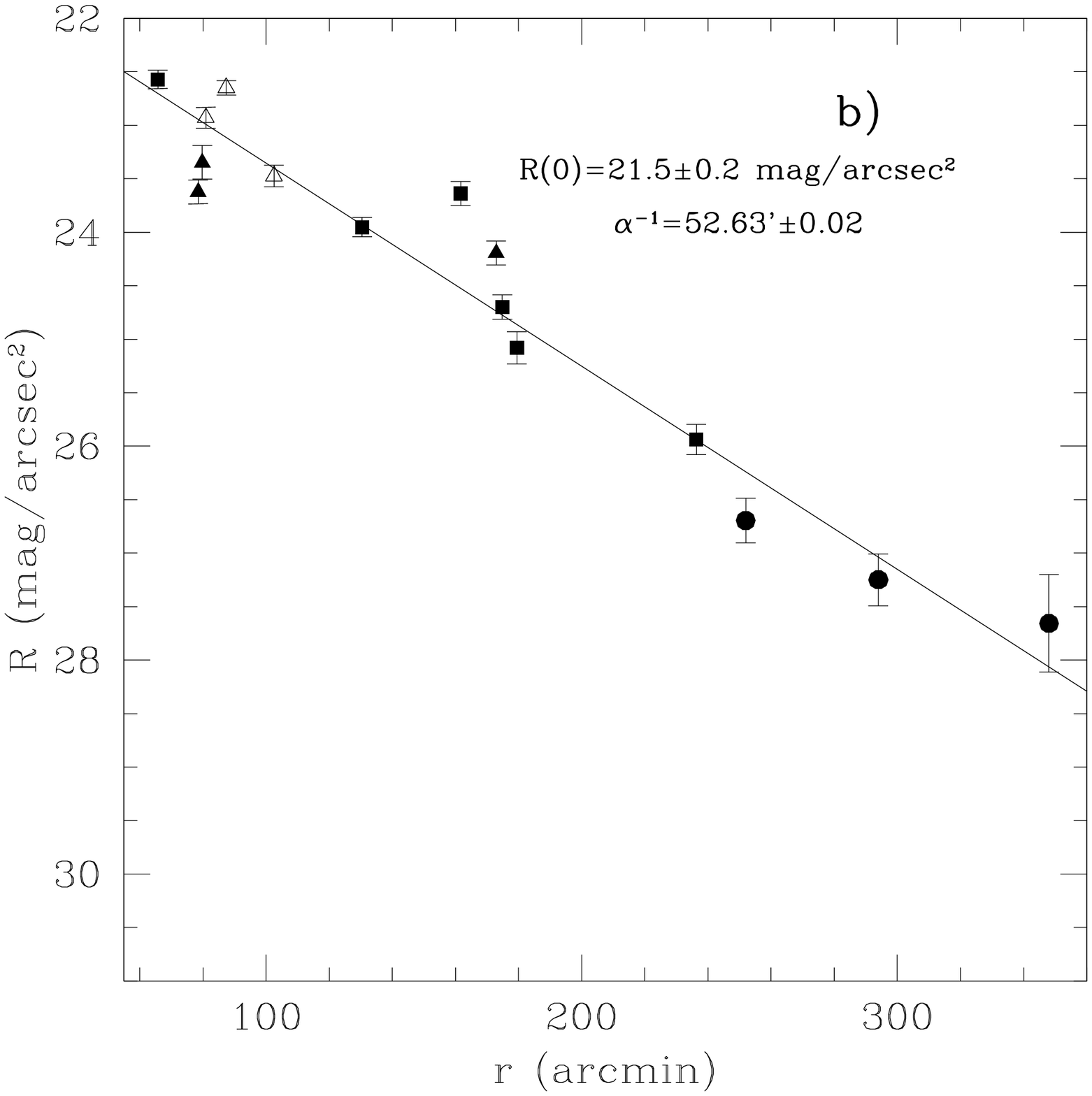}
\caption{The surface brightness profile of the SMC in the ${\it B}$ (panel a) and ${\it R}$ (panel b) bands. The filled circles
 represent the three outer SMC fields discussed in this paper. Open triangles represent eastern SMC fields while
 filled triangles denote the western SMC fields (qj0033-7028, qj0036-7225, qj0037-7218, qj0111-7249, qj0112-7236,
  qj0116-7259 in figure~\ref{SMC})
   and the squares denote the southern fields (smc0057-7353, smc0100-7457, smc0049-7544 in figure~\ref{SMC})
    taken from No\"el et al. 2007. The profile has been fitted with an exponential disk (solid line).}
\label{surface_brightness}
\end{figure}

In this paper, we have shown that, at 6.5 kpc from the SMC center, there is still a stellar population
belonging to the SMC, with a substantial amount of intermediate-age stars.
 Fields located at 4.7 kpc and 5.6 kpc show stars as young as 3 Gyr old
 and the field at 6.5 kpc as young as 5 Gyr old. 
   From the analysis of the CMDs, no strong
population gradients are present in the outer SMC disk, from 
$\thicksim$2.7\degr outwards (field qj0047-7530 in figure~\ref{SMC}, see
No\"el et al. 2007).
 The origin of the intermediate-age population in the outer SMC fields presented 
here is still uncertain. If these stars formed at their current positions, it would imply that the SMC is significantly more
extended (and therefore more massive) than previously thought. Its original gas envelope must have 
  extended much further away in the past. 
 
 Alternatively, these stars could represent tidal debris torn off through
interactions with the Milky Way or the LMC. 
However, the surface brightness distribution in the outer parts is quite smooth.
 Detailed
models of tidal interactions show flat or rising surface brightness distributions (e.g. Johnston et al. 2002; Read et al. 2006b),
 unlike the falling surface 
brightness profile we find here. In fact, the 
surface brightness profile is well fit by an exponential 
disk, indicating that the SMC disk is dominating over a possible old Milky Way-like 
halo.
 Our results indicate that the tidal radius of the SMC disk
is larger than the 5 kpc proposed by Gardiner \& Noguchi (1996) and the 6.3 kpc recently suggested by Connors et al. (2006).
This strengthens the hypothesis of a larger size for the SMC.
 An upper limit for the SMC size in the direction of the Magellanic Bridge is provided by Harris (2007), who finds no signs of an
 intermediate-age or old population (but only young Bridge population) at  6.4\degr from the center. Note, however, that Harris' field is
 located along the ``minor axis'' of the main body and therefore  could be misleading to do a direct comparison. 
   More studies further away from the SMC and with large spatial coverage are needed to confirm the 
actual extent of the SMC.

\acknowledgments

We would like to thank Justin I. Read for very useful comments and a careful reading of the manuscript in its early stages.
 We are grateful for helpful discussions with Antonio Aparicio. 
 N.E.D.N. would like to thank Matteo Monelli for kindly supplying his IDL programs and his experience with WFI data.   
We thank the anonymous referee for valuable comments. 
The authors acknowledge support by the Plan Nacional de Investigaci\'on Cient\'ifica, Desarrollo, e Investigaci\'on Tecnol\'ogica,
(AYA2004-06343).

\clearpage

\begin{deluxetable}{rrrrrrrrr}
\tabletypesize{\scriptsize}
\tablewidth{0pt}
\setlength{\tabcolsep}{0.05in}
\tablecaption{Data obtained from the outskirts of the SMC}
\tablehead{
\colhead{Field} & \colhead{$\alpha_{2000}$} & \colhead{$\delta_{2000}$} & r({\degr})\tablenotemark{a} & \colhead{{\it B}-band exposures (sec)} &
\colhead{{\it R}-band exposures (sec)} }
\startdata

F1 & 00:53:00 & -77:00:00 & 4.2 &21$\times$407.9+2$\times$89.9 & 17$\times$256.9+2$\times$89.9\\  
F2 & 00:52:52 & -77:46:00 & 4.9 &12$\times$407.9+1$\times$89.9&  15$\times$256.9$\times$89.9\\ 
F3 & 00:52:40 & -78:34:00 & 5.8 &13$\times$407.9 & 12$\times$256.9+1$\times$89.9\\

\enddata
\tablenotetext{a}{Distance from the SMC center, $\alpha_{2000}=$ 00:52:44.8, $\delta_{2000}=$ -72:49:43}
\label{fields}
\end{deluxetable}

\clearpage  

\clearpage

\clearpage

\end{document}